\documentclass[12pt,a4]{article}
\usepackage{amsmath, graphicx}
\usepackage{amstext,amsfonts,amsbsy,eucal,amssymb, amsthm}
\usepackage{color}
\newtheorem{theorem}{Theorem}[section]

\theoremstyle{definition}

\newtheorem{remark}[theorem]{Remark}

\def\C{\operatorname{\mathbb{C}}}

\usepackage{float}
\restylefloat{table}
\usepackage{graphicx}
\DeclareSymbolFont{extraup}{U}{zavm}{m}{n}
\DeclareMathSymbol{\varheart}{\mathalpha}{extraup}{86}
\DeclareMathSymbol{\vardiamond}{\mathalpha}{extraup}{87}

\newcommand{\tr}{\mathrm{tr}}

\DeclareFontFamily{U}{rsf}{}
\DeclareFontShape{U}{rsf}{m}{n}{<5> <6> rsfs5 <7> <8> <9> rsfs7 <10-> rsfs10}{}
\DeclareMathAlphabet\Scr{U}{rsf}{m}{n}
\def\C{{\mathbb{C}}}

\def\tr{\operatorname{{\rm tr}}}
\def\str{\operatorname{{\rm str}}}
\def\det{\operatorname{{\rm det}}}
\def\sdet{\operatorname{{\rm sdet}}}

\usepackage{graphics}
\usepackage{epsfig}
\usepackage{amsmath}
\usepackage{amssymb}
\usepackage{multirow}
\usepackage[all]{xy}

\begin{document}
\begin{center}
{\bf Super-QRT and 4D-mappings reduced from the lattice super-KdV equation}
\end{center}
\begin{center}
A. S. Carstea, T. Takenawa, 
\end{center}
\begin{center}
{\it National Institute of Physics and Nuclear Engineering, Dept. of Theoretical Physics, Atomistilor 407, 077125, Magurele, Bucharest, Romania}\\ 
{\it Faculty of Marine Technology, Tokyo University of Marine Science  and Technology, 2-1-6 Etchu-jima, Koto-ku, Tokyo, 135-8533, Japan}\\
\end{center}

\begin{abstract}
Starting from the complete integrable lattice super-KdV equation,
two super-mappings are obtained by performing a travelling-wave reduction. The first one is linear and the second is a four dimensional super-QRT mapping
containing both Grassmann commuting and anti-commuting dependent variables. Adapting the classical ``staircase'' method to the Lax super-matrices of the lattice super-KdV equation, we compute the Lax super-matrices of the mapping and the two invariants; the first one is a pure nilpotent commuting quantity and the second one is given by an elliptic curve containing nilpotent commuting Grassmann coefficients as well.
In the case of finitely generated Grassmann algebra with two generators, the super-QRT mapping becomes a four-dimensional ordinary discrete dynamical system that has two invariants but does not satisfy singularity confinement criterion. It is also observed that the dynamical degree of this system grows quadratically.
\end{abstract}

\section{Introduction}

The QRT mapping (the name coming from the initials of the authors, Quispel, Roberts and Thomson) \cite{qrt} is a fundamental object in the theory of integrable systems and discrete Painlev\'e equations. It is an order two autonomous mapping depending on 16 constant parameters, with the solution given by the values of an elliptic function at equidistant points on a line in a complex torus. On the other hand, it is at the heart of any lattice soliton equation, since the travelling-wave reduction of them is usually a QRT map \cite{red}. In algebraic geometry it is also a very interesting object representing the ``dynamical'' equivalent of the famous {\it group law} of an elliptic curve \cite{tsuda}, realizing in fact an automorphism of the pencil of the elliptic curves parametrised by its integral \cite{tsuda},\cite{dui}. 

As we mention before, the QRT mapping is at the heart of the construction of discrete Painlev\'e equations \cite{basil-book}. More precisely, they 
can be regarded as integrable deautonomized QRT mappings based on the so called ``singularity confinement''\cite{sc}. In  
the discrete setting a singularity of a second order mapping arises at the value of the variable, say $x_n$,  
such that $x_{n+1}$ does not depend on $x_{n-1}$ and $x_n$ as independent variables. If this situation disappears after a finite number of iterations and the values of $x_{n-1}$ and $x_n$ are recovered (memory of initial condition is not lost unlike the chaotic dynamics) then we have the {\it singularity confinement} phenomenon. 
It is a  useful detector of integrability (however not sufficient) and the QRT mapping satisfies this criterion in its full generality. Moreover it was very productive and a big number of discrete Painlev\'e equations have been obtained \cite{basil-book}. Their integrable character has been settled definitively by finding Lax pairs, bilinear forms, B\"acklund/Schlesinger transformations, special solutions etc.\cite{basilbook2-b}.

In the present paper we consider an example of a fermionic extension of a QRT mapping. It is a coupled system of nonlinear discrete equations having two dependent variables with values in the commutative (bosonic) and anti-commutative (fermionic) sector of an infinite dimensional Grassmann algebra. The motivation comes from the recent construction of lattice super-KdV equation \cite{liu}, where the Lax pair, consistency around the cube and super-multisoliton solution were constructed \cite{fane}. In this paper we 
consider the traveling wave reduction of the lattice super-KdV which gives an example of a super-QRT mapping as a fourth order mapping. Using the ``staircase method''\cite{frank}, we will construct its Lax operators as a pair of even super-matrices. The super-trace of the spectral Lax super-matrix gives two invariants, one being fully nilpotent and the other one a biquadratic expression containing Grassmann even coefficients.

In the case of a finitely generated Grassmann algebra with two generators,
our super-QRT mapping splits into a four-dimensional ordinary discrete dynamical system possessing two invariants. Different from the autonomous case of discrete four dimensional Painlev\'e equations previously studied by the authors \cite{CT2018}, the present system is a coupling of a QRT map and a linear map whose coefficients are functions of the solution of the former. Using generating function technique\cite{HV97}, we observe that the dynamical degrees of this system and a reduced three dimensional system grow quadratically, while previously known such coupling systems have cubic degree growth \cite{Lafortune2001, Gubbiotti2018}. 
A rigorous proof requires detailed study on resolution of indeterminacy for the defining manifold, and will be given in a subsequent report. 

This paper is organised as follows. 
In Section 2 we recall the lattice super-symmetric KdV equation and its Lax formalism.
In Section 3 a super-QRT mapping is deduced by traveling-wave reduction.
Applying the staircase method for the Lax system, we also obtain its invariants. 
In Section 4 we deduce a four dimensional classical discrete dynamical system in the case of a finitely generated Grassmann algebra with two generators. The growth of the dynamical degrees of this system and a reduced three dimensional system is studied.

\section{Lattice super-KdV equation and the staircase method}
In this section we shall work with lattice super-symmetric KdV equation \cite{liu}:
\begin{align}\label{diliu1}
&\psi_{n+1,m+1}-\psi_{n,m}=\frac{2(p_1+p_2)(\psi_{n+1,m}-\psi_{n,m+1})}{2(p_2-p_1)+u_{n+1,m}-u_{n,m+1}}\\
&u_{n+1,m+1}-u_{n,m}=\frac{2(p_1+p_2)(u_{n+1,m}-u_{n,m+1})}{2(p_2-p_1)+u_{n+1,m}-u_{n,m+1}} \nonumber\\ \label{diliu2}
&-\frac{(p_1+p_2)(4(p_2-p_1)+u_{n+1,m}-u_{n,m+1})}{(2(p_2-p_1)+u_{n+1,m}-u_{n,m+1})^2}(\psi_{n+1,m}-\psi_{n,m+1})(\psi_{n,m+1}-\psi_{n,m}),
\end{align}
where $u_{n,m}$ and $\psi_{n,m}$ are functions with values in the commuting (bosonic) and anti-commuting (fermionic) sector of an infinite dimensional Grassmann algebra, 
i.e. $u_{n,m}u_{n',m'}=u_{n',m'}u_{n,m}$, $\psi_{n,m}\psi_{n',m'}=-\psi_{n',m'}\psi_{n,m}, u_{n,m}\psi_{n,m}=\psi_{n,m}u_{n,m}$; also $p_1, p_2$ are parameters of the lattice which for simplicity we consider to be ordinary complex numbers. 
Recall that a Grassmann algebra $L$ is decomposed as $L=L_{{\rm even}}\oplus L_{{\rm odd}}$
such that $uv = -v u$ if $u, v \in L_{{\rm odd}}$ and $uv = v u$ otherwise;
$L_{{\rm even}}^2\subset L_{{\rm even}}$, $L_{{\rm odd}}^2\subset L_{{\rm even}}$   
and $L_{{\rm odd}} L_{{\rm even}}=L_{{\rm even}} L_{{\rm odd}} \subset L_{{\rm odd}}$. The subspaces $L_{{\rm even}}$ and $L_{{\rm odd}}$ are called the commuting and the anti-commuting sectors respectively.    
This system is given by the compatibility condition:
\begin{align}
&A_{n,m+1}B_{n,m}-B_{n+1,m}A_{n,m}=0
\end{align}
for the associated spectral problems:
\begin{align}\label{m}
\Phi_{n,m+1}=&B_{n,m}\Phi_{n,m},\quad \Phi_{n+1,m}=A_{n,m}\Phi_{n,m},
\end{align}
where
\begin{align*}
A_{n,m}=
\left(
\begin{array}{cccc}
p_1&1&-\eta&0\\
\lambda^2&p_1&-2p_1\eta+\eta g&-\eta\\
0&\eta&p_1-g&1\\
\lambda^2 \eta&2p_1\eta-\eta g&\lambda^2-2p_1g+g^2&p_1-g\\
\end{array}\right)
\end{align*}
\begin{align*}
B_{n,m}=
\left(
\begin{array}{cccc}
p_2&1&-\sigma&0\\
\lambda^2&p_2&-2p_2\sigma+\sigma f&-\sigma\\
0&\sigma&p_2-f&1\\
\lambda^2 \sigma&2p_2\sigma-\sigma f&\lambda^2-2p_2f+f^2&p_2-f\\
\end{array}\right)
\end{align*}
with 
$$\eta=\frac{1}{2}(\psi_{n+1,m}-\psi_{n,m}),\quad g=\frac{1}{2}(u_{n+1,m}-u_{n,m})$$
$$\sigma=\frac{1}{2}(\psi_{n,m+1}-\psi_{n,m}),\quad f=\frac{1}{2}(u_{n,m+1}-u_{n,m})$$
and $\lambda$ is the spectral parameter which is a commuting invertible number (Grassmann commuting number with nonzero body).

The super-matrices $A_{n,m}$ and $B_{n,m}$ are even-super-matrices i.e. they are made by four $2\times 2$ blocks, the blocks on the diagonal being formed by Grassmann commuting (bosonic)  numbers and the off-diagonal blocks are made of Grassmann anti-commuting (fermionic) numbers \cite{s}. For any even super-matrix
\begin{align*}
 M=\left(
 \begin{array}{cc}
  P&R\\
  T&Q\\
 \end{array}\right),
\end{align*}
one can define the super-trace which is the difference of ordinary traces of blocks $P$ and $Q$
$$\str M=\tr P-\tr Q,$$
and the super-determinant
$$\sdet M=\det (P-RQ^{-1}T)(\det Q)^{-1}.$$ 
A very important property of the super-trace is the following; for any super-matrices $M$, $N$ we have \cite{s}
\begin{equation}\label{str}
 \str (MN)=\str (NM).
\end{equation}
Moreover, if the block submatrices $P$ and $Q$ are nonsingular, then the super-determinant has nonzero body and the super-matrix $M$ is invertible. In our case both the super-matrices $A_{n,m}$ and $B_{n,m}$ are even invertible super-matrices.

In the $(p,q)$ traveling-wave reduction (or $(p,q)$ periodic reduction) we consider that our dependent variables are periodic with period $(p,q)$ and will depend only on one variable $\nu=p m- q n$, and accordingly, our system will turn into a system of nonlinear ordinary discrete super-equations.
This reduction modifies the structure of Lax compatibility, since we have only one independent variable. To see how we get the Lax pairs for the reduced system, we implement the well known ``staircase method'' \cite{frank}.
The method relies on the fact that $(p,q)$ traveling reduction perform a periodic problem on the staircase in the $(n,m)$ lattice (the lattice system being quadrilateral and the initial conditions defined on a {\it stair}). In this periodic problem, one of the Lax operators is transformed into a monodromy matrix ($L_{\nu}$) and the other one ($M_{\nu}$) will be related to evolution on $\nu$. These new Lax even super-matrices are given by the following relations \cite{chris}:
\begin{equation}\label{st1}
L_{\nu}=\sideset{}{'}\prod_{j=0}^{q-1}B_{n+p,m+j}\sideset{}{'}\prod_{i=0}^{p-1}A_{n+i,m}
\end{equation}
\begin{equation}\label{st2}
M_{\nu}=\sideset{}{'}\prod_{j=0}^{c-1}B_{n+c,m+j}\sideset{}{'}\prod_{i=0}^{d-1}A_{n+i,m},
\end{equation}
where $\sideset{}{'}\prod$ follows the order of the stair (in our subsequent case the order will be just reversible order) and $(c,d)$ are related to the 
shift $\nu\to \nu+1$ in the way $(n,m)\to(n+c,m+d)$ \cite{chris}. However in our applications we will consider the simplest case $c=d=1$.

Now the compatibility condition $A_{n,m+1}B_{n,m}-B_{n+1,m}A_{n,m}=0$ will turn into $L_{\nu+1}M_{\nu}-M_{\nu}L_{\nu}=0$

\begin{remark}: 
The computations given in \cite{chris} can be translated one to one to the case of even supermatrices.

\end{remark}

\section{Super-QRT mapping}

Let us consider the simplest choice, the travelling-wave reduction 
$(p,q)=(2,1)$, namely $\nu=2m-n$. So in this case $u_{n+1,m}\to u_{\nu-1},u_{n+1,m+1}\to u_{\nu+1}$, etc. Our system will turn into:

\begin{align}
&\psi_{\nu+1}-\psi_{\nu}=\frac{2(p_1+p_2)(\psi_{\nu-1}-\psi_{\nu+2})}{2(p_2-p_1)+u_{\nu-1}-u_{\nu+2}}\\
&u_{\nu+1}-u_{\nu}=\frac{2(p_1+p_2)(u_{\nu-1}-u_{\nu+2})}{2(p_2-p_1)+u_{\nu-1}-u_{\nu+2}}\nonumber\\
&-\frac{(p_1+p_2)(4(p_2-p_1)+u_{\nu-1}-u_{\nu+2})}{(2(p_2-p_1)+u_{\nu-1}-u_{\nu+2})^2}(\psi_{\nu-1}-\psi_{\nu+2})(\psi_{\nu+2}-\psi_{\nu}).
\end{align}
This system is of  order six. We can reduce the order by ``integrating'' once 
each equation in the system. Defining  $x_{\nu}=u_{\nu+1}-u_{\nu}$  and  $\zeta_{\nu}=\psi_{\nu+1}-\psi_{\nu}$,
we obtain the 4D system:
\begin{align}
&\zeta_{\nu}=\frac{2(p_1+p_2)(-\zeta_{\nu-1}-\zeta_{\nu}-\zeta_{\nu+1})}{2(p_2-p_1)-x_{\nu-1}-x_{\nu}-x_{\nu+1}}\\
&x_{\nu}=\frac{2(p_1+p_2)(-x_{\nu-1}-x_{\nu}-x_{\nu+1})}{2(p_2-p_1)-x_{\nu-1}-x_{\nu}-x_{\nu+1}} \nonumber \\ \label{s2}
&+(p_1+p_2)(\zeta_{\nu-1}+\zeta_{\nu}+\zeta_{\nu+1})(\zeta_{\nu+1}+\zeta_{\nu})\frac{4(p_2-p_1)-x_{\nu-1}-x_{\nu}-x_{\nu+1}}{(2(p_2-p_1)-x_{\nu-1}-x_{\nu}-x_{\nu+1})^2},
\end{align}
where $x_{\mu}$ is a commuting variable and  $\zeta_{\mu}$ is an anti-commuting variable.

Solving for $x_{\nu-1}+x_{\nu}+x_{\nu+1}$, we 
obtain two coupled systems corresponding to the roots:
\begin{align}\label{4th-1-1}
\zeta_{\nu-1}+\zeta_{\nu}+\zeta_{\nu+1}=&\left(\frac{x_{\nu-1}+x_{\nu}+x_{\nu+1}}{2(p_1+p_2)}-\frac{p_2-p_1}{p_2+p_1}\right)\zeta_{\nu}\\\label{4th-1-2}
x_{\nu-1}+x_{\nu}+x_{\nu+1}=&\frac{2x_{\nu}(p_1-p_2)}{2(p_2+p_1)-x_{\nu}}+\frac{x_{\nu}-4p_1-4p_2}{2x_{\nu}-4p_1-4p_2}\zeta_{\nu-1}(\zeta_{\nu}+\zeta_{\nu+1})
\end{align}
and
\begin{align}\label{4th-2-1}
\zeta_{\nu-1}+\zeta_{\nu}+\zeta_{\nu+1}=&\left(\frac{x_{\nu-1}+x_{\nu}+x_{\nu+1}}{2(p_1+p_2)}-\frac{p_2-p_1}{p_2+p_1}\right)\zeta_{\nu}\\ \label{4th-2-2}
x_{\nu-1}+x_{\nu}+x_{\nu+1}=&-2(p_1-p_2)-\frac{1}{2}\zeta_{\nu-1}(\zeta_{\nu}+\zeta_{\nu+1}).
\end{align}
Here, we used the fact that the equation (\ref{s2}) is quadratic and because of the nilpotent character of the Grassmann zeta's, the square root of the discriminant disappears  
through the formula $\sqrt{a+b\zeta_1\zeta_2}=\sqrt{a}(1+\frac{1}{2}\zeta_1\zeta_2 b)$.

Let us study  the system \eqref{4th-2-1}, \eqref{4th-2-2}. 
Solving the second equation for $x_{\nu-1}+x_{\nu}+x_{\nu+1}$ and substituting it into the first one, we get the following simpler form:
\begin{align}\label{linn}
\zeta_{\nu-1}+\zeta_{\nu}+\zeta_{\nu+1}=&\frac{\zeta_{\nu-1}\zeta_{\nu}\zeta_{\nu+1}}{4(p_2+p_1)}\\\label{linn1}
x_{\nu-1}+x_{\nu}+x_{\nu+1}=&-2(p_1-p_2)-\frac{1}{2}\zeta_{\nu-1}(\zeta_{\nu}+\zeta_{\nu+1}).
\end{align}
But for any three Grassmann nilpotent numbers $a,b,c$ with the property $a+b+c=\alpha abc$ (with $\alpha\in \C$), we have $a+b+c=0$. Indeed, $a(a+b+c)=a(\alpha abc)=\alpha a^2bc=0$ (because $a^2=b^2=c^2=0$ from nilpotency) implies $ab+ac=0$. Accordingly $abc=(ab)c=(-ac)c=-ac^2=0$. Putting $a=\zeta_{\nu-1}, b=\zeta_{\nu}, c=\zeta_{\nu+1}$, we will have $\zeta_{\nu-1}+\zeta_{\nu}+\zeta_{\nu+1}=\zeta_{\nu-1}(\zeta_{\nu}+\zeta_{\nu+1})=0$, and  so the system (\ref{linn}), (\ref{linn1}) is a trivial linear uncoupled one:
\begin{align}
\zeta_{\nu-1}+\zeta_{\nu}+\zeta_{\nu+1}=&0\\
x_{\nu-1}+x_{\nu}+x_{\nu+1}=&-2(p_1-p_2).
\end{align}

Next let us study the first one \eqref{4th-1-1}, \eqref{4th-1-2}. One can see immediately that this system is a super-QRT map (for $\zeta_{\nu}=0$ it is reduced to an ordinary QRT map). However we can further simplify it;
multiplying  $\zeta_{\nu}$ to the first equation from the left, 
we obtain $\zeta_{\nu}\zeta_{\nu+1}+\zeta_{\nu}\zeta_{\nu-1}=0$, and hence
\begin{align}\label{4th-1-1b}\zeta_{\nu}\zeta_{\nu+1}= \zeta_{\nu-1}\zeta_{\nu}.\end{align}
Also, using
\begin{align*}
\zeta_{\nu-1}(\zeta_{\nu}+\zeta_{\nu+1})=&(\zeta_{\nu-1}+\zeta_{\nu}+\zeta_{\nu+1})(\zeta_{\nu}+\zeta_{\nu+1})\\
=&\left(\frac{x_{\nu-1}+x_{\nu}+x_{\nu+1}}{2(p_1+p_2)}-\frac{p_2-p_1}{p_2+p_1}\right)\zeta_{\nu}\zeta_{\nu+1}
\end{align*}
and replacing $x_{\nu-1}+x_{\nu}+x_{\nu+1}$ by \eqref{4th-1-2}, 
\eqref{4th-1-2} becomes
\begin{align}
x_{\nu-1}+x_{\nu}+x_{\nu+1}=\frac{2x_{\nu}(p_1-p_2)}{2(p_2+p_1)-x_{\nu}}+(p_1-p_2)\frac{4p_1+4p_2-x_{\nu}}{(2p_1+2p_2-x_{\nu})^2}\zeta_{\nu}\zeta_{\nu+1}.
\end{align}

The equation \eqref{4th-1-1b} implies $\gamma_{\nu}=\zeta_{\nu}\zeta_{\nu+1}$ to be a constant of motion($\gamma_{\nu+1}=\gamma_{\nu}$ for all $\nu$), which is 
the {\it first conservation law} and we denote it by $\gamma$.  
The constant $\gamma$ is a nilpotent ($\gamma^2=0$) commuting Grassmanian number (depending of course on the initial condition for the odd Grassmann variable 
$\zeta_0\zeta_1$). Hence, our system 
can  be reduced to a single 
second order mapping depending on $\gamma$:
\begin{align}\label{2nd-1}
x_{\nu-1}+x_{\nu}+x_{\nu+1}=\frac{2x_{\nu}(p_1-p_2)}{2(p_2+p_1)-x_{\nu}}+(p_1-p_2)\frac{4p_1+4p_2-x_{\nu}}{(2p_1+2p_2-x_{\nu})^2}\gamma.
\end{align}

For finding the second conserved quantity, we have to use the Lax pair. Applying the staircase method to the Lax pair of lattice super-KdV equation (relations (\ref{st1}), (\ref{st2})), we 
have the following Lax operators for the system \eqref{4th-1-1}, \eqref{4th-1-2}: 
$$L_\nu=B_{\nu-2} A_{\nu-1} A_{\nu},\quad M_{\nu}=B_{\nu-1}A_{\nu}$$
where $A_{\nu}$ and $B_{\nu}$ are given by (\ref{m}) but depending on the combination $\nu=2m-n$ 
through
\begin{align*}
 &A_{\nu}=
 \left(
\begin{array}{cccc}
p_1&1&-\alpha_1&0\\
\lambda^2&p_1&-\beta_1&-\alpha_1\\
0&\alpha_1&a_1&1\\
\lambda^2\alpha_1 &\beta_1&b_1&a_1\\
\end{array}\right),\quad
 B_{\nu}=
 \left(
\begin{array}{cccc}
p_2&1&-\alpha_2&0\\
\lambda^2&p_2&-\beta_2&-\alpha_2\\
0&\alpha_2&a_2&1\\
\lambda^2\alpha_2 &\beta_2&b_2&a_2\\
\end{array}\right).
\end{align*}
with
\begin{align*}
\alpha_1=&\frac{\zeta_{\nu-1}}{2}\\
\beta_1=&p_1\zeta_{\nu-1}+\frac{1}{4}x_{\nu-1}\zeta_{\nu-1}\\
a_1=&p_1+\frac{x_{\nu-1}}{2}\\
b_1=&\lambda^2-p_1(x_{\nu}+x_{\nu+1})+\frac{1}{4}(x_{\nu}+x_{\nu+1})^2
\end{align*}
and
\begin{align*}
\alpha_2=&\frac{1}{2}(\zeta_{\nu}+\zeta_{\nu+1})\\
\beta_2=&\frac{1}{2}p_2(\zeta_{\nu}+\zeta_{\nu+1})-\frac{1}{4}(x_{\nu}+x_{\nu+1})(\zeta_{\nu}+\zeta_{\nu+1})\\
a_2=&p_2-\frac{1}{2}(x_{\nu}+x_{\nu+1})\\
b_2=&\lambda^2-p_2(x_{\nu}+x_{\nu+1})+\frac{1}{4}(x_{\nu}+x_{\nu+1})^2.
\end{align*}
The compatibility condition of the Lax supermatrices $L_{\nu}$ and $M_{\nu}$ is $L_{\nu+1}M_{\nu}=M_{\nu}L_{\nu}\Leftrightarrow L_{\nu+1}=M_{\nu}L_{\nu}M_{\nu}^{-1}$. So applying the supertrace we have:
$$\str L_{\nu+1}=\str (M_{\nu}L_{\nu}M_{\nu}^{-1}).$$ 
From the property (\ref{str}) we have $\str(M_{\nu}L_{\nu}M_{\nu}^{-1})=\str(L_{\nu}M_{\nu}^{-1}M_{\nu})=\str L_{\nu}$. So the supertrace of $L_{\nu}$ is invariant at the evolution $\nu\to \nu+1$. Expanding str$(L_{\nu})$ in powers of the spectral parameter the corresponding coefficients will be the conservation laws. After long, but straightforward computations we obtain the second conserved quantity:
\begin{align}\nonumber
I=&\frac{1}{8}(2p_1(x_{\nu}^2 + x_{\nu-1}^2)+(x_{\nu} + x_{\nu-1}) (-x_{\nu} x_{\nu-1} + 2p_2(x_{\nu} + x_{\nu-1})))\\& 
+  \gamma (p_1(p_1+2p_2) - \frac{1}{16} ( (4 p_1 - 4 p_2 + x_{\nu}) (x_{\nu}+x_{\nu-1}) + x_{\nu-1}^2)).
\end{align}
 Further, by the following rescalings, 
$$x_{\nu}\to 2 x_{\nu} (p_1 + p_2),\quad \gamma\to 4\gamma \frac{(p_1 + p_2)^2}{(p_1 - p_2)},\quad p_2\to p_1\frac{ (1 - h)}{(1 + h)}$$  
the equation \eqref{2nd-1} is simplified as:
\begin{align}\label{qrt}
x_{\nu+1}+x_{\nu}+x_{\nu-1}=&\frac{hx_{\nu}}{1-x_{\nu}}+\frac{2-x_{\nu}}{(1-x_{\nu})^2}\gamma,
\end{align}
while the invariant is simplified as
\begin{align}\label{inv}
I(x,y)=&h(x^2(y - 1)-y^2+xy(y+h-1))+\gamma(x^2+xy+y^2+2h(x + y)),
\end{align}
where $x_{\nu}\equiv x, x_{\nu-1}\equiv y$ are Grassmann commuting function, and $h$ is an ordinary complex number.
One can see that the this invariant is expressed by an elliptic curve with coefficients in the commuting sector of the Grassmann algebra.

\begin{remark}
Because $\gamma$ is a commuting number and $\gamma^2=0$, we can consider the term $(2-x_{\nu})\gamma/(1-x_{\nu})^2$ in (\ref{qrt})
as if it is  {\it a first order perturbation} of the ordinary QRT mapping
$$ x_{\nu+1}+x_{\nu}+x_{\nu-1}=\frac{hx_{\nu}}{1-x_{\nu}}$$
($\gamma$ being here a small 
parameter) and the invariant (\ref{inv}) to be the ordinary QRT invariant plus the order-one perturbative correction (and the conservation law is valid only in the first order, namely $I(x_{\nu+1},y_{\nu+1})-I(x_{\nu},y_{\nu})=O(\gamma^2)$). In this way, the integrable Grassmann extensions can be useful for the perturbation theory of integrable mappings.

On the other hand it seems that the singularities plays a weaker role in the case of super-mappings. Indeed the first integral is not depending at all on the right hand side of the first (pure fermionic) equation. So for any equation
$$\zeta_{\nu+1}+\zeta_{\nu}+\zeta_{\nu-1}=G(x_\nu)\zeta_{\nu}$$
we do have the integral $I=\zeta_{\nu}\zeta_{\nu+1}$ no matter how bad the singularities of $G(x_{\nu})$ are. In addition, the first integral \eqref{inv} is not useful to reduce the system to lower dimensional one, since we can not solve algebraic equation including nilpotent Grassmann numbers.

\end{remark}

\section{Finitely generated Grassmann algebra}

In the case of a finitely generated Grassmann algebra, we can rewrite our system \eqref{qrt} into a higher dimensional ordinary discrete dynamical system. 
We assume that the Grassmann algebra is generated by two generators $\{\xi_1, \xi_2 \}$. In this base, $x_{\nu}=x_{\nu}^{(0)}+ x_{\nu}^{(3)} \xi_1 \xi_2 $ (with ordinary complex functions $x_\nu^{(i)}$'s) and, for simplicity, we consider $\gamma=\xi_1\xi_2$, $h$ an ordinary complex parameter. Then we have
$$
(x_{\nu+1}^{(0)}+ x_{\nu}^{(0)}+ x_{\nu-1}^{(0)})+ 
(x_{\nu+1}^{(3)}+ x_{\nu}^{(3)}+ x_{\nu-1}^{(3)})\gamma
=
\frac{hx_{\nu}^{(0)}}{1-x_{\nu}^{(0)}}
+ \frac{2-x_{\nu}^{(0)}+hx_{\nu}^{(3)}}{(1-x_{\nu}^{(0)})^2}\gamma.
$$
Setting as $x_0= x_{\nu-1}^{(0)}$,  $x_1= x_{\nu-1}^{(3)}$, $x_2 = x_{\nu}^{(0)} $, $x_3= x_{\nu}^{(3)}$ and $\bar{x}_0= x_{\nu}^{(0)}$,  $\bar{x}_1= x_{\nu}^{(3)}$, $\bar{x}_2 = x_{\nu+1}^{(0)} $, $\bar{x}_3= x_{\nu+1}^{(3)}$, this equation becomes a four dimensional system as
\begin{align}\label{4d}
\varphi:&\left\{\begin{array}{rcl}
\overline{x_0}&=&x_2\\
\overline{x_1}&=&x_3\\
\overline{x_2}&=&-x_2-x_0+\dfrac{h x_2}{1-x_2}\\
\overline{x_3}&=&-x_1-x_3 + \dfrac{2-x_2+hx_3}{(1-x_2)^2}
\end{array}\right..
\end{align}
This system is a QRT map for variables $x_0$, $x_2$ coupled with linear equations for variables $x_1$, $x_3$ with coefficients depending on $x_2$.
In general, for $N$ generators, we will obtain a $2N$-dimensional system consisting in a full nonlinear QRT
map coupled with linear equations.

The invariant (\ref{inv}) is also decomposed and we get the following two invariants for the system (\ref{4d}):
\begin{align}
I_1=&-hx_0^2 - hx_0x_2 + h^2 x_0 x_2 + h x_0^2 x_2 - h x_2^2 + h x_0 x_2^2\\
\nonumber
I_2=&2 h x_0 + x_0^2 - 2 h x_0 x_1 + 2 h x_2 + x_0 x_2 - h x_1 x_2 + h^2 x_1 x_2
+2 h x_0 x_1 x_2\\& + x_2^2
 + h x_1 x_2^2 - h x_0 x_3 + h^2 x_0 x_3 + h x_0^2 x_3 - 
  2 h x_2 x_3 + 2 h x_0 x_2 x_3.
\end{align}

For a birational map $\varphi:\C^N\to \C^N$, the growth of the polynomial degree of (numerator or denominator of) the $n$-iterate ($\varphi^n$) with respect to the initial coordinates is an important indicator for its integrability or chaosity, i.e. the former may occur if the degree grows polynomially, while the latter occurs if it grows exponentially. 

Let $\varphi$ be given by $(\ref{4d})$ and denote the $n$-iterate $\varphi^n(x_0,x_1,x_2,x_3)$ as $(x_0^{(n)},x_1^{(n)},x_2^{(n)},x_3^{(n)})$.  Then, the degree of $x_3^{(n)}$ with respect to $x_2$ is 
$$0, 2, 6, 10, 16, 24, 32, 42, 54, 66, 80, \dots.$$
The above numbers are observed (not proved) to be the coefficients of the series of the following generating function (cf. \cite{HV97}):
$$ g(s)=\frac{-2 s ( s + 1)}{(s^2 + s + 1) (s - 1)^3}=2s+6s^2+10s^3+16s^4+24s^5+32s^6+\dots.$$
Due to the presence of $(s-1)^3$ in the denominator, the growth is quadratic in accord with the fact that the mapping is integrable.

The singularity confinement criterion proposed by Grammaticos-Ramani and their collaborators \cite{sc, rgh} is also
an important indicator for its integrability. Let us start with a hyper-plane $x_1 = \varepsilon^{-1}$, where 
$\varepsilon $ is a small parameter for considering Laurent series expression, and apply $\varphi$ and $\varphi^{-1}$, then  we have a ``unconfined'' sequence of Laurent series
\begin{align}
\cdots \to \mbox{ 2-dim subvariety} \to&\left( \left(-1+ \frac{h}{(x_0^{(0)}-1)^2}\right) \varepsilon^{-1},x_1^{(-1)},x_2^{(-1)},\varepsilon^{-1}\right)\nonumber\\
\to& 
(x_0^{(0)},\varepsilon^{-1},x_2^{(0)},x_3^{(0)})
\to 
(x_2^{(0)},x_3^{(0)},x_2^{(1)},\varepsilon^{-1})\nonumber\\
\to &
\left(x_2^{(1)},\varepsilon^{-1}, x_2^{(2)} , \left(-1+\frac{h}{(x_2^{(0)}-1)^{2}}\right)\varepsilon^{-1}\right)\nonumber\\
\to& \mbox{ 2-dim subvariety} \to \cdots,\label{seq3}
\end{align}
where $x_i^{(k)}$'s are complex constants and only the principal term is written for each entry.

Since the degree of the conserved quantity $I_2$ with respect to $x_1$ or $x_3$ is one, we can restrict the phase space of the system $(\ref{4d})$ into a 3-dimensional space and the system is thus reduced to a 3-dimensional one:
\begin{align}\label{3d_system}
\psi:&\left\{
\begin{array}{rcl}
x_0&=&x_2\\ 
x_1&=& \frac{I_2 - (x_0^2+x_0x_2+x_2^2)-2h(x_0-x_0x_1+x_2)-hx_1x_2(2x_0+x_2-1+h)}{h(- x_0  +  h x_0 +  x_0^2 - 2  x_2 + 2  x_0 x_2) }\\
x_2&=& -x_2 - x_0 + \frac{h x_2}{1-x_2}
\end{array}\right..
\end{align}
The growth of $x_1^{(n)}$ with respect to $x_0$ is $0, 2, 2, 4, 8, 12, 18, 26, 34, 44, \dots$.
and its generating function is observed to be
$$ g(s)=\frac{-2s (s^4 + s^2- s+1)}{(s^2 + s + 1) (s - 1)^3}.$$
Due to the presence of $(s-1)^3$ in the denominator, the growth is quadratic.

As reported in \cite{Lafortune2001} or recently in \cite{Gubbiotti2018}, it is well-known that 
a coupling of  an integrable mapping of quadratic degree growth with a linear mapping has cubic degree growth.
Our four dimensional system is also a coupling of an integrable mapping of quadratic degree growth with a linear mapping but has only quadratic degree growth. Moreover, its reduction to three dimensional system has also quadratic degree growth.

The technique we used  above for computing dynamical degree growth is not rigorous one.
In a subsequent paper, we will give an algebraic variety where Systems \eqref{4d} is lifted to
be algebraically stable mappings \cite{CT2018} and, using actions on the Picard group, prove their degree growth actually to be quadratic.\\

\section{Conclusions}

In this article we studied the travelling-wave reduction of the integrable lattice supersymmetric KdV equation. Because of the Grassmann character of the dependent variables we obtained two maps, one proved to be linear and the other one a super-QRT map. Using the staircase method, the Lax super-matrices of the lattice super-KdV were reduced to the Lax super-matrices of the super-QRT map. This fact allowed computation of the conserved quantities, one a bodyless commuting one and the other expressed by a biquadratic expression with Grassmann even coefficients. The interesting fact is that one can use the finitely generated Grassmann algebra to obtain various multidimensional ordinary discrete integrable dynamical systems. In our case we performed the analysis for the case of a Grassmann algebra with two generators and found a four dimensional map possessing two invariants. This system is a coupling of a QRT map and a liner map and does not satisfy the singularity confinement property. Its complexity growth turned out to be only quadratic, different from previously known such coupled systems.

\subsection*{Acknowledgement}
ASC was supported by Program Nuccleu, PN/2019, Romanian Ministery of Education
and T.~T. was supported by the Japan Society for the
Promotion of Science, Grand-in-Aid (C) (17K05271).

\end{document}